\documentclass[conference]{IEEEtran}
\IEEEoverridecommandlockouts
\usepackage{cite}
\usepackage{amsmath,amssymb,amsfonts}
\usepackage{algorithm}
\usepackage{algorithmic}
\usepackage{graphicx}
\usepackage{textcomp}
\usepackage{xcolor}
\usepackage{hyperref}
\hypersetup{
    colorlinks=true,
    urlcolor=[RGB]{25,168,255},
    citecolor=[RGB]{5,92,41},
    linkcolor=[RGB]{5,92,41},
    linkbordercolor = {white}
}
\usepackage[draft]{minted}

\makeatletter
\def\PYG@reset{\let\PYG@it=\relax \let\PYG@bf=\relax%
    \let\PYG@ul=\relax \let\PYG@tc=\relax%
    \let\PYG@bc=\relax \let\PYG@ff=\relax}
\def\PYG@tok#1{\csname PYG@tok@#1\endcsname}
\def\PYG@toks#1+{\ifx\relax#1\empty\else%
    \PYG@tok{#1}\expandafter\PYG@toks\fi}
\def\PYG@do#1{\PYG@bc{\PYG@tc{\PYG@ul{%
    \PYG@it{\PYG@bf{\PYG@ff{#1}}}}}}}
\def\PYG#1#2{\PYG@reset\PYG@toks#1+\relax+\PYG@do{#2}}

\@namedef{PYG@tok@w}{\def\PYG@tc##1{\textcolor[rgb]{0.73,0.73,0.73}{##1}}}
\@namedef{PYG@tok@c}{\let\PYG@it=\textit\def\PYG@tc##1{\textcolor[rgb]{0.24,0.48,0.48}{##1}}}
\@namedef{PYG@tok@cp}{\def\PYG@tc##1{\textcolor[rgb]{0.61,0.40,0.00}{##1}}}
\@namedef{PYG@tok@k}{\let\PYG@bf=\textbf\def\PYG@tc##1{\textcolor[rgb]{0.00,0.50,0.00}{##1}}}
\@namedef{PYG@tok@kp}{\def\PYG@tc##1{\textcolor[rgb]{0.00,0.50,0.00}{##1}}}
\@namedef{PYG@tok@kt}{\def\PYG@tc##1{\textcolor[rgb]{0.69,0.00,0.25}{##1}}}
\@namedef{PYG@tok@o}{\def\PYG@tc##1{\textcolor[rgb]{0.40,0.40,0.40}{##1}}}
\@namedef{PYG@tok@ow}{\let\PYG@bf=\textbf\def\PYG@tc##1{\textcolor[rgb]{0.67,0.13,1.00}{##1}}}
\@namedef{PYG@tok@nb}{\def\PYG@tc##1{\textcolor[rgb]{0.00,0.50,0.00}{##1}}}
\@namedef{PYG@tok@nf}{\def\PYG@tc##1{\textcolor[rgb]{0.00,0.00,1.00}{##1}}}
\@namedef{PYG@tok@nc}{\let\PYG@bf=\textbf\def\PYG@tc##1{\textcolor[rgb]{0.00,0.00,1.00}{##1}}}
\@namedef{PYG@tok@nn}{\let\PYG@bf=\textbf\def\PYG@tc##1{\textcolor[rgb]{0.00,0.00,1.00}{##1}}}
\@namedef{PYG@tok@ne}{\let\PYG@bf=\textbf\def\PYG@tc##1{\textcolor[rgb]{0.80,0.25,0.22}{##1}}}
\@namedef{PYG@tok@nv}{\def\PYG@tc##1{\textcolor[rgb]{0.10,0.09,0.49}{##1}}}
\@namedef{PYG@tok@no}{\def\PYG@tc##1{\textcolor[rgb]{0.53,0.00,0.00}{##1}}}
\@namedef{PYG@tok@nl}{\def\PYG@tc##1{\textcolor[rgb]{0.46,0.46,0.00}{##1}}}
\@namedef{PYG@tok@ni}{\let\PYG@bf=\textbf\def\PYG@tc##1{\textcolor[rgb]{0.44,0.44,0.44}{##1}}}
\@namedef{PYG@tok@na}{\def\PYG@tc##1{\textcolor[rgb]{0.41,0.47,0.13}{##1}}}
\@namedef{PYG@tok@nt}{\let\PYG@bf=\textbf\def\PYG@tc##1{\textcolor[rgb]{0.00,0.50,0.00}{##1}}}
\@namedef{PYG@tok@nd}{\def\PYG@tc##1{\textcolor[rgb]{0.67,0.13,1.00}{##1}}}
\@namedef{PYG@tok@s}{\def\PYG@tc##1{\textcolor[rgb]{0.73,0.13,0.13}{##1}}}
\@namedef{PYG@tok@sd}{\let\PYG@it=\textit\def\PYG@tc##1{\textcolor[rgb]{0.73,0.13,0.13}{##1}}}
\@namedef{PYG@tok@si}{\let\PYG@bf=\textbf\def\PYG@tc##1{\textcolor[rgb]{0.64,0.35,0.47}{##1}}}
\@namedef{PYG@tok@se}{\let\PYG@bf=\textbf\def\PYG@tc##1{\textcolor[rgb]{0.67,0.36,0.12}{##1}}}
\@namedef{PYG@tok@sr}{\def\PYG@tc##1{\textcolor[rgb]{0.64,0.35,0.47}{##1}}}
\@namedef{PYG@tok@ss}{\def\PYG@tc##1{\textcolor[rgb]{0.10,0.09,0.49}{##1}}}
\@namedef{PYG@tok@sx}{\def\PYG@tc##1{\textcolor[rgb]{0.00,0.50,0.00}{##1}}}
\@namedef{PYG@tok@m}{\def\PYG@tc##1{\textcolor[rgb]{0.40,0.40,0.40}{##1}}}
\@namedef{PYG@tok@gh}{\let\PYG@bf=\textbf\def\PYG@tc##1{\textcolor[rgb]{0.00,0.00,0.50}{##1}}}
\@namedef{PYG@tok@gu}{\let\PYG@bf=\textbf\def\PYG@tc##1{\textcolor[rgb]{0.50,0.00,0.50}{##1}}}
\@namedef{PYG@tok@gd}{\def\PYG@tc##1{\textcolor[rgb]{0.63,0.00,0.00}{##1}}}
\@namedef{PYG@tok@gi}{\def\PYG@tc##1{\textcolor[rgb]{0.00,0.52,0.00}{##1}}}
\@namedef{PYG@tok@gr}{\def\PYG@tc##1{\textcolor[rgb]{0.89,0.00,0.00}{##1}}}
\@namedef{PYG@tok@ge}{\let\PYG@it=\textit}
\@namedef{PYG@tok@gs}{\let\PYG@bf=\textbf}
\@namedef{PYG@tok@gp}{\let\PYG@bf=\textbf\def\PYG@tc##1{\textcolor[rgb]{0.00,0.00,0.50}{##1}}}
\@namedef{PYG@tok@go}{\def\PYG@tc##1{\textcolor[rgb]{0.44,0.44,0.44}{##1}}}
\@namedef{PYG@tok@gt}{\def\PYG@tc##1{\textcolor[rgb]{0.00,0.27,0.87}{##1}}}
\@namedef{PYG@tok@err}{\def\PYG@bc##1{{\setlength{\fboxsep}{\string -\fboxrule}\fcolorbox[rgb]{1.00,0.00,0.00}{1,1,1}{\strut ##1}}}}
\@namedef{PYG@tok@kc}{\let\PYG@bf=\textbf\def\PYG@tc##1{\textcolor[rgb]{0.00,0.50,0.00}{##1}}}
\@namedef{PYG@tok@kd}{\let\PYG@bf=\textbf\def\PYG@tc##1{\textcolor[rgb]{0.00,0.50,0.00}{##1}}}
\@namedef{PYG@tok@kn}{\let\PYG@bf=\textbf\def\PYG@tc##1{\textcolor[rgb]{0.00,0.50,0.00}{##1}}}
\@namedef{PYG@tok@kr}{\let\PYG@bf=\textbf\def\PYG@tc##1{\textcolor[rgb]{0.00,0.50,0.00}{##1}}}
\@namedef{PYG@tok@bp}{\def\PYG@tc##1{\textcolor[rgb]{0.00,0.50,0.00}{##1}}}
\@namedef{PYG@tok@fm}{\def\PYG@tc##1{\textcolor[rgb]{0.00,0.00,1.00}{##1}}}
\@namedef{PYG@tok@vc}{\def\PYG@tc##1{\textcolor[rgb]{0.10,0.09,0.49}{##1}}}
\@namedef{PYG@tok@vg}{\def\PYG@tc##1{\textcolor[rgb]{0.10,0.09,0.49}{##1}}}
\@namedef{PYG@tok@vi}{\def\PYG@tc##1{\textcolor[rgb]{0.10,0.09,0.49}{##1}}}
\@namedef{PYG@tok@vm}{\def\PYG@tc##1{\textcolor[rgb]{0.10,0.09,0.49}{##1}}}
\@namedef{PYG@tok@sa}{\def\PYG@tc##1{\textcolor[rgb]{0.73,0.13,0.13}{##1}}}
\@namedef{PYG@tok@sb}{\def\PYG@tc##1{\textcolor[rgb]{0.73,0.13,0.13}{##1}}}
\@namedef{PYG@tok@sc}{\def\PYG@tc##1{\textcolor[rgb]{0.73,0.13,0.13}{##1}}}
\@namedef{PYG@tok@dl}{\def\PYG@tc##1{\textcolor[rgb]{0.73,0.13,0.13}{##1}}}
\@namedef{PYG@tok@s2}{\def\PYG@tc##1{\textcolor[rgb]{0.73,0.13,0.13}{##1}}}
\@namedef{PYG@tok@sh}{\def\PYG@tc##1{\textcolor[rgb]{0.73,0.13,0.13}{##1}}}
\@namedef{PYG@tok@s1}{\def\PYG@tc##1{\textcolor[rgb]{0.73,0.13,0.13}{##1}}}
\@namedef{PYG@tok@mb}{\def\PYG@tc##1{\textcolor[rgb]{0.40,0.40,0.40}{##1}}}
\@namedef{PYG@tok@mf}{\def\PYG@tc##1{\textcolor[rgb]{0.40,0.40,0.40}{##1}}}
\@namedef{PYG@tok@mh}{\def\PYG@tc##1{\textcolor[rgb]{0.40,0.40,0.40}{##1}}}
\@namedef{PYG@tok@mi}{\def\PYG@tc##1{\textcolor[rgb]{0.40,0.40,0.40}{##1}}}
\@namedef{PYG@tok@il}{\def\PYG@tc##1{\textcolor[rgb]{0.40,0.40,0.40}{##1}}}
\@namedef{PYG@tok@mo}{\def\PYG@tc##1{\textcolor[rgb]{0.40,0.40,0.40}{##1}}}
\@namedef{PYG@tok@ch}{\let\PYG@it=\textit\def\PYG@tc##1{\textcolor[rgb]{0.24,0.48,0.48}{##1}}}
\@namedef{PYG@tok@cm}{\let\PYG@it=\textit\def\PYG@tc##1{\textcolor[rgb]{0.24,0.48,0.48}{##1}}}
\@namedef{PYG@tok@cpf}{\let\PYG@it=\textit\def\PYG@tc##1{\textcolor[rgb]{0.24,0.48,0.48}{##1}}}
\@namedef{PYG@tok@c1}{\let\PYG@it=\textit\def\PYG@tc##1{\textcolor[rgb]{0.24,0.48,0.48}{##1}}}
\@namedef{PYG@tok@cs}{\let\PYG@it=\textit\def\PYG@tc##1{\textcolor[rgb]{0.24,0.48,0.48}{##1}}}

\def\PYGZus{\char`\_}
\def\PYGZob{\char`\{}
\def\PYGZcb{\char`\}}

\def\PYGZsh{\char`\#}

\def\PYGZhy{\char`\-}

\def\PYGZdq{\char`\"}

\makeatother

\makeatletter
\def\PYGdefault@reset{\let\PYGdefault@it=\relax \let\PYGdefault@bf=\relax%
    \let\PYGdefault@ul=\relax \let\PYGdefault@tc=\relax%
    \let\PYGdefault@bc=\relax \let\PYGdefault@ff=\relax}
\def\PYGdefault@tok#1{\csname PYGdefault@tok@#1\endcsname}
\def\PYGdefault@toks#1+{\ifx\relax#1\empty\else%
    \PYGdefault@tok{#1}\expandafter\PYGdefault@toks\fi}
\def\PYGdefault@do#1{\PYGdefault@bc{\PYGdefault@tc{\PYGdefault@ul{%
    \PYGdefault@it{\PYGdefault@bf{\PYGdefault@ff{#1}}}}}}}
\def\PYGdefault#1#2{\PYGdefault@reset\PYGdefault@toks#1+\relax+\PYGdefault@do{#2}}

\@namedef{PYGdefault@tok@w}{\def\PYGdefault@tc##1{\textcolor[rgb]{0.73,0.73,0.73}{##1}}}
\@namedef{PYGdefault@tok@c}{\let\PYGdefault@it=\textit\def\PYGdefault@tc##1{\textcolor[rgb]{0.24,0.48,0.48}{##1}}}
\@namedef{PYGdefault@tok@cp}{\def\PYGdefault@tc##1{\textcolor[rgb]{0.61,0.40,0.00}{##1}}}
\@namedef{PYGdefault@tok@k}{\let\PYGdefault@bf=\textbf\def\PYGdefault@tc##1{\textcolor[rgb]{0.00,0.50,0.00}{##1}}}
\@namedef{PYGdefault@tok@kp}{\def\PYGdefault@tc##1{\textcolor[rgb]{0.00,0.50,0.00}{##1}}}
\@namedef{PYGdefault@tok@kt}{\def\PYGdefault@tc##1{\textcolor[rgb]{0.69,0.00,0.25}{##1}}}
\@namedef{PYGdefault@tok@o}{\def\PYGdefault@tc##1{\textcolor[rgb]{0.40,0.40,0.40}{##1}}}
\@namedef{PYGdefault@tok@ow}{\let\PYGdefault@bf=\textbf\def\PYGdefault@tc##1{\textcolor[rgb]{0.67,0.13,1.00}{##1}}}
\@namedef{PYGdefault@tok@nb}{\def\PYGdefault@tc##1{\textcolor[rgb]{0.00,0.50,0.00}{##1}}}
\@namedef{PYGdefault@tok@nf}{\def\PYGdefault@tc##1{\textcolor[rgb]{0.00,0.00,1.00}{##1}}}
\@namedef{PYGdefault@tok@nc}{\let\PYGdefault@bf=\textbf\def\PYGdefault@tc##1{\textcolor[rgb]{0.00,0.00,1.00}{##1}}}
\@namedef{PYGdefault@tok@nn}{\let\PYGdefault@bf=\textbf\def\PYGdefault@tc##1{\textcolor[rgb]{0.00,0.00,1.00}{##1}}}
\@namedef{PYGdefault@tok@ne}{\let\PYGdefault@bf=\textbf\def\PYGdefault@tc##1{\textcolor[rgb]{0.80,0.25,0.22}{##1}}}
\@namedef{PYGdefault@tok@nv}{\def\PYGdefault@tc##1{\textcolor[rgb]{0.10,0.09,0.49}{##1}}}
\@namedef{PYGdefault@tok@no}{\def\PYGdefault@tc##1{\textcolor[rgb]{0.53,0.00,0.00}{##1}}}
\@namedef{PYGdefault@tok@nl}{\def\PYGdefault@tc##1{\textcolor[rgb]{0.46,0.46,0.00}{##1}}}
\@namedef{PYGdefault@tok@ni}{\let\PYGdefault@bf=\textbf\def\PYGdefault@tc##1{\textcolor[rgb]{0.44,0.44,0.44}{##1}}}
\@namedef{PYGdefault@tok@na}{\def\PYGdefault@tc##1{\textcolor[rgb]{0.41,0.47,0.13}{##1}}}
\@namedef{PYGdefault@tok@nt}{\let\PYGdefault@bf=\textbf\def\PYGdefault@tc##1{\textcolor[rgb]{0.00,0.50,0.00}{##1}}}
\@namedef{PYGdefault@tok@nd}{\def\PYGdefault@tc##1{\textcolor[rgb]{0.67,0.13,1.00}{##1}}}
\@namedef{PYGdefault@tok@s}{\def\PYGdefault@tc##1{\textcolor[rgb]{0.73,0.13,0.13}{##1}}}
\@namedef{PYGdefault@tok@sd}{\let\PYGdefault@it=\textit\def\PYGdefault@tc##1{\textcolor[rgb]{0.73,0.13,0.13}{##1}}}
\@namedef{PYGdefault@tok@si}{\let\PYGdefault@bf=\textbf\def\PYGdefault@tc##1{\textcolor[rgb]{0.64,0.35,0.47}{##1}}}
\@namedef{PYGdefault@tok@se}{\let\PYGdefault@bf=\textbf\def\PYGdefault@tc##1{\textcolor[rgb]{0.67,0.36,0.12}{##1}}}
\@namedef{PYGdefault@tok@sr}{\def\PYGdefault@tc##1{\textcolor[rgb]{0.64,0.35,0.47}{##1}}}
\@namedef{PYGdefault@tok@ss}{\def\PYGdefault@tc##1{\textcolor[rgb]{0.10,0.09,0.49}{##1}}}
\@namedef{PYGdefault@tok@sx}{\def\PYGdefault@tc##1{\textcolor[rgb]{0.00,0.50,0.00}{##1}}}
\@namedef{PYGdefault@tok@m}{\def\PYGdefault@tc##1{\textcolor[rgb]{0.40,0.40,0.40}{##1}}}
\@namedef{PYGdefault@tok@gh}{\let\PYGdefault@bf=\textbf\def\PYGdefault@tc##1{\textcolor[rgb]{0.00,0.00,0.50}{##1}}}
\@namedef{PYGdefault@tok@gu}{\let\PYGdefault@bf=\textbf\def\PYGdefault@tc##1{\textcolor[rgb]{0.50,0.00,0.50}{##1}}}
\@namedef{PYGdefault@tok@gd}{\def\PYGdefault@tc##1{\textcolor[rgb]{0.63,0.00,0.00}{##1}}}
\@namedef{PYGdefault@tok@gi}{\def\PYGdefault@tc##1{\textcolor[rgb]{0.00,0.52,0.00}{##1}}}
\@namedef{PYGdefault@tok@gr}{\def\PYGdefault@tc##1{\textcolor[rgb]{0.89,0.00,0.00}{##1}}}
\@namedef{PYGdefault@tok@ge}{\let\PYGdefault@it=\textit}
\@namedef{PYGdefault@tok@gs}{\let\PYGdefault@bf=\textbf}
\@namedef{PYGdefault@tok@gp}{\let\PYGdefault@bf=\textbf\def\PYGdefault@tc##1{\textcolor[rgb]{0.00,0.00,0.50}{##1}}}
\@namedef{PYGdefault@tok@go}{\def\PYGdefault@tc##1{\textcolor[rgb]{0.44,0.44,0.44}{##1}}}
\@namedef{PYGdefault@tok@gt}{\def\PYGdefault@tc##1{\textcolor[rgb]{0.00,0.27,0.87}{##1}}}
\@namedef{PYGdefault@tok@err}{\def\PYGdefault@bc##1{{\setlength{\fboxsep}{\string -\fboxrule}\fcolorbox[rgb]{1.00,0.00,0.00}{1,1,1}{\strut ##1}}}}
\@namedef{PYGdefault@tok@kc}{\let\PYGdefault@bf=\textbf\def\PYGdefault@tc##1{\textcolor[rgb]{0.00,0.50,0.00}{##1}}}
\@namedef{PYGdefault@tok@kd}{\let\PYGdefault@bf=\textbf\def\PYGdefault@tc##1{\textcolor[rgb]{0.00,0.50,0.00}{##1}}}
\@namedef{PYGdefault@tok@kn}{\let\PYGdefault@bf=\textbf\def\PYGdefault@tc##1{\textcolor[rgb]{0.00,0.50,0.00}{##1}}}
\@namedef{PYGdefault@tok@kr}{\let\PYGdefault@bf=\textbf\def\PYGdefault@tc##1{\textcolor[rgb]{0.00,0.50,0.00}{##1}}}
\@namedef{PYGdefault@tok@bp}{\def\PYGdefault@tc##1{\textcolor[rgb]{0.00,0.50,0.00}{##1}}}
\@namedef{PYGdefault@tok@fm}{\def\PYGdefault@tc##1{\textcolor[rgb]{0.00,0.00,1.00}{##1}}}
\@namedef{PYGdefault@tok@vc}{\def\PYGdefault@tc##1{\textcolor[rgb]{0.10,0.09,0.49}{##1}}}
\@namedef{PYGdefault@tok@vg}{\def\PYGdefault@tc##1{\textcolor[rgb]{0.10,0.09,0.49}{##1}}}
\@namedef{PYGdefault@tok@vi}{\def\PYGdefault@tc##1{\textcolor[rgb]{0.10,0.09,0.49}{##1}}}
\@namedef{PYGdefault@tok@vm}{\def\PYGdefault@tc##1{\textcolor[rgb]{0.10,0.09,0.49}{##1}}}
\@namedef{PYGdefault@tok@sa}{\def\PYGdefault@tc##1{\textcolor[rgb]{0.73,0.13,0.13}{##1}}}
\@namedef{PYGdefault@tok@sb}{\def\PYGdefault@tc##1{\textcolor[rgb]{0.73,0.13,0.13}{##1}}}
\@namedef{PYGdefault@tok@sc}{\def\PYGdefault@tc##1{\textcolor[rgb]{0.73,0.13,0.13}{##1}}}
\@namedef{PYGdefault@tok@dl}{\def\PYGdefault@tc##1{\textcolor[rgb]{0.73,0.13,0.13}{##1}}}
\@namedef{PYGdefault@tok@s2}{\def\PYGdefault@tc##1{\textcolor[rgb]{0.73,0.13,0.13}{##1}}}
\@namedef{PYGdefault@tok@sh}{\def\PYGdefault@tc##1{\textcolor[rgb]{0.73,0.13,0.13}{##1}}}
\@namedef{PYGdefault@tok@s1}{\def\PYGdefault@tc##1{\textcolor[rgb]{0.73,0.13,0.13}{##1}}}
\@namedef{PYGdefault@tok@mb}{\def\PYGdefault@tc##1{\textcolor[rgb]{0.40,0.40,0.40}{##1}}}
\@namedef{PYGdefault@tok@mf}{\def\PYGdefault@tc##1{\textcolor[rgb]{0.40,0.40,0.40}{##1}}}
\@namedef{PYGdefault@tok@mh}{\def\PYGdefault@tc##1{\textcolor[rgb]{0.40,0.40,0.40}{##1}}}
\@namedef{PYGdefault@tok@mi}{\def\PYGdefault@tc##1{\textcolor[rgb]{0.40,0.40,0.40}{##1}}}
\@namedef{PYGdefault@tok@il}{\def\PYGdefault@tc##1{\textcolor[rgb]{0.40,0.40,0.40}{##1}}}
\@namedef{PYGdefault@tok@mo}{\def\PYGdefault@tc##1{\textcolor[rgb]{0.40,0.40,0.40}{##1}}}
\@namedef{PYGdefault@tok@ch}{\let\PYGdefault@it=\textit\def\PYGdefault@tc##1{\textcolor[rgb]{0.24,0.48,0.48}{##1}}}
\@namedef{PYGdefault@tok@cm}{\let\PYGdefault@it=\textit\def\PYGdefault@tc##1{\textcolor[rgb]{0.24,0.48,0.48}{##1}}}
\@namedef{PYGdefault@tok@cpf}{\let\PYGdefault@it=\textit\def\PYGdefault@tc##1{\textcolor[rgb]{0.24,0.48,0.48}{##1}}}
\@namedef{PYGdefault@tok@c1}{\let\PYGdefault@it=\textit\def\PYGdefault@tc##1{\textcolor[rgb]{0.24,0.48,0.48}{##1}}}
\@namedef{PYGdefault@tok@cs}{\let\PYGdefault@it=\textit\def\PYGdefault@tc##1{\textcolor[rgb]{0.24,0.48,0.48}{##1}}}

\makeatother

\newcommand{\angstrom}{\mbox{\normalfont\AA}}

\begin{document}

\title{Extending OpenKIM with an Uncertainty Quantification Toolkit for Molecular Modeling}

\author{
    \IEEEauthorblockN{Yonatan Kurniawan}
    \IEEEauthorblockA{\textit{Department of Physics and Astronomy} \\
	\textit{Brigham Young University}\\
	Provo, UT, USA \\
	yonatank@byu.edu, ORCID 0000-0002-5369-5710}\\
    \IEEEauthorblockN{Mark K. Transtrum}
    \IEEEauthorblockA{\textit{Department of Physics and Astronomy} \\
	\textit{Brigham Young University}\\
	Provo, UT, USA \\
	mktranstrum@byu.edu, ORCID 0000-0001-9529-9399}\\
    \IEEEauthorblockN{Ryan S. Elliott}
    \IEEEauthorblockA{\textit{Department of Aerospace Engineering and Mechanics} \\
	\textit{University of Minnesota}\\
	Minneapolis, MN, USA\\
	relliott@umn.edu, ORCID 0000-0003-4988-8306}
	\and
    \IEEEauthorblockN{Cody L. Petrie}
    \IEEEauthorblockA{\textit{Department of Physics and Astronomy} \\
	\textit{Brigham Young University}\\
	Provo, UT, USA \\
	codypetrie89@gmail.com, ORCID 0000-0002-5596-8516}\\
    \IEEEauthorblockN{Ellad B. Tadmor}
    \IEEEauthorblockA{\textit{Department of Aerospace Engineering and Mechanics} \\
	\textit{University of Minnesota}\\
	Minneapolis, MN, USA \\
	tadmor@umn.edu, ORCID 0000-0003-3311-6299}\\
    \IEEEauthorblockN{Daniel S. Karls}
    \IEEEauthorblockA{\textit{Department of Aerospace Engineering and Mechanics} \\
	\textit{University of Minnesota}\\
	Minneapolis, MN, USA \\
	karl0100@umn.edu, ORCID 0000-0002-4069-396X}
    \and
    \IEEEauthorblockN{Mingjian Wen}
    \IEEEauthorblockA{\textit{Energy Technologies Area} \\
	\textit{Lawrence Berkeley National Laboratory}\\
	Berkeley, CA, USA \\
	mjwen@lbl.gov, ORCID 0000-0003-0013-575X}
}

\maketitle

\begin{abstract}
    Atomistic simulations are an important tool in materials modeling.
    Interatomic potentials (IPs) are at the heart of such molecular models, and the accuracy of a model's predictions depends strongly on the choice of IP.
    Uncertainty quantification (UQ) is an emerging tool for assessing the reliability of atomistic simulations.
    The Open Knowledgebase of Interatomic Models (OpenKIM) is a cyberinfrastructure project whose goal is to collect and standardize the study of IPs to enable transparent, reproducible research.
    Part of the OpenKIM framework is the Python package, KIM-based Learning-Integrated Fitting Framework (KLIFF), that provides tools for fitting parameters in an IP to data.
    This paper introduces a UQ toolbox extension to KLIFF.
    We focus on two sources of uncertainty: variations in parameters and inadequacy of the functional form of the IP.
    Our implementation uses parallel-tempered Markov chain Monte Carlo (PTMCMC), adjusting the sampling temperature to estimate the uncertainty due to the functional form of the IP.
    We demonstrate on a Stillinger--Weber potential that makes predictions for the atomic energies and forces for silicon in a diamond configuration.
    Finally, we highlight some potential subtleties in applying and using these tools with recommendations for practitioners and IP developers.
\end{abstract}


\section{Introduction}
\label{sec:introduction}
Molecular modeling is an important part of materials science, and interatomic potentials (IPs) are at the heart of most molecular modeling simulations\cite{Brenner_2000}.
Given the large number of IPs constructed for various applications, there is an acute need to standardize their computational implementation and evaluation and facilitate portability among researchers.
To this end, the Open Knowledgebase of Interatomic Models (OpenKIM) was founded to make atomistic-scale simulations reliable, reproducible, and accessible\cite{Tadmor_Elliott_Sethna_Miller_Becker_2011, Roadmap_MSMSE_2020}.
While the OpenKIM framework provides many tools for materials simulations, it does not provide a standardized tool for uncertainty quantification (UQ).
In this paper, we describe a novel UQ toolbox within the OpenKIM framework, targeted at molecular modeling.

% Introduction to molecular modeling
The general goal of molecular modeling is to predict properties of materials by simulating collections of atoms.
In this setting, IPs are used to approximate the interaction energy of atoms as functions of the atomic positions and species\cite{LeSar2013-cp}.
IPs are used in conjunction with a simulation program, such as ASE\cite{ase_paper} or LAMMPS\cite{LAMMPS_paper}, to model atomic behavior and extract material properties.
Such simulations can be static, e.g., minimizing energy to obtain the equilibrium lattice parameter of a crystal, or dynamic, e.g. applying the fluctuation--dissipation theorem to compute thermal conductivity.
The accuracy of the material properties predicted by an atomistic simulation depends strongly on the choice of IP, and considerable effort has gone into developing IPs that are accurate for specific applications.

UQ is an emerging field of applied mathematics that aims to quantify and reduce uncertainties in mathematical models\cite{Kennedy_OHagan_2001}.
In molecular modeling, UQ can help assess the reliability of conclusions drawn from atomistic simulations.
It is widely recognized that the largest source of uncertainty in molecular modeling is the functional form of the IP.
Additional uncertainty comes from the values of the corresponding parameters, which are typically fit to experimental data or first-principles (quantum mechanical) calculations \cite{Ercolessi_Adams_1994}.
Having been fit to data, these IPs are often used to calculate out-of-sample material properties or to conduct large scale simulations.
These simulations generally suffer from inconsistent transferability, i.e., they may struggle to accurately predict material properties to which the IPs are not fit\cite{Karls}.
The UQ process is especially important for assessing the reliability of these out-of-sample predictions.

There are many UQ methods that have been used in molecular modeling, including $F$-statistics estimations \cite{Messerly_Knotts_Wilding_2017}, ANOVA-based methods \cite{Tschopp_Chris_Rinderspacher_Nouranian_Baskes_Gwaltney_Horstemeyer_2018}, multi-objective optimization \cite{Mishra_Hong_Rajak_Sheng_Nomura_Kalia_Nakano_Vashishta_2018}, and profile likelihoods\cite{kurniawan2021bayesian}.
Among these methods, the most frequently utilized is a Bayesian method known as Markov chain Monte Carlo (MCMC)\cite{Frederiksen_Jacobsen_Brown_Sethna_2004, Angelikopoulos_Papadimitriou_Koumoutsakos_2012, Rizzi_Najm_Debusschere_Sargsyan_Salloum_Adalsteinsson_Knio_2012, Rizzi_Jones_Debusschere_Knio_2013, Cools-Ceuppens_Verstraelen_2017, Messerly_Shirts_Kazakov_2018, De_Simon_Iglesias_Jones_Wood_2018, Vohra_Nobakht_Shin_Mahadevan_2018, Vohra_Mahadevan_2019, Dhaliwal_Nair_Singh_2019a, Dhaliwal_Nair_Singh_2019b, Longbottom_Brommer_2019}.
As such, we focus on MCMC methods, though we anticipate future extensions that will make other tools available to the molecular modeling community.

A universal challenge in applying UQ methods to multi-parameter models, such as IPs, is that models are often \emph{sloppy} \cite{Wen_Li_Brommer_Elliott_Sethna_Tadmor_2016, Wen_Shirodkar_Plechas_Kaxiras_Elliott_Tadmor_2017, kurniawan2021bayesian}.
Sloppiness refers to an extremely ill-conditioned inference problem when fitting parameters to data, which is ubiquitous in many scientific fields 
\cite{Jeong_Zhuang_Transtrum_Zhou_Qiu_2018, White_Tolman_Thames_Withers_Mason_Transtrum_2016, Transtrum_Qiu_2016, Mannakee_Ragsdale_Transtrum_Gutenkunst_2016,  Transtrum_Machta_Brown_Daniels_Myers_Sethna_2015, Transtrum_Qiu_2012, transtrum2017information, Gutenkunst_2007, Machta_Chachra_Transtrum_Sethna_2013, Waterfall_Casey_Gutenkunst_Brown_Myers_Brouwer_Elser_Sethna_2006}.
For sloppy models, many parameter combinations are not well-constrained by available data and dubbed \emph{practically unidentifiable}.
This poses a number of challenges to clearly formulating and interpreting UQ results, as shown, for example, for molecular modeling \cite{parameterinflation, kurniawan2021bayesian}.
These subtleties need to be considered, as we demonstrate later in this paper, when performing UQ analyses.

Although many software or libraries for performing Bayesian sampling or other UQ methods exist, such as emcee\cite{Foreman-Mackey_Hogg_Lang_Goodman_2013}, Chaospy\cite{chaospy_paper}, and EasyVVUQ\cite{easyvvuq_paper} to name a few, but there are only few libraries that integrate UQ for molecular modeling, such as \textit{potfit}\cite{potfit_paper, Longbottom_Brommer_2019}.
In this paper we introduce a UQ toolkit integrated within the OpenKIM framework to facilitate the application of UQ to molecular modeling.
Integrating UQ capabilities directly into the OpenKIM framework allows for more uniform, reproducible results and helps to standardize the practice of reporting uncertainty as part of a molecular modeling workflow.
In Sec.~\ref{sec:background}, we introduce the theory behind UQ and describe in more detail the OpenKIM framework.
We describe our specific UQ implementation in Sec.~\ref{sec:solution} and how to use it, with an example, in Sec.~\ref{sec:results}.
Finally, we conclude and describe future directions for this UQ toolkit in Sec.~\ref{sec:conclusion}.

\section{Background}
\label{sec:background}

\subsection{Theory of Uncertainty Quantification}
\label{subsec:probability_uq_theory}

Uncertainty in modeling may have many sources such as stochasticity in data or numerical discretization.
In molecular modeling, the dominant source of uncertainty is due to the functional form of the IP, sometimes known as \emph{model inadequacy}.
That is to say, the IP does not capture all of the physics present in the process it is intended to mimic.
An IP is meant to encompass some effects of quantum mechanics, but it necessarily does not include all quantum effects, leading to uncertainty when the IP predicts material properties different from those on which it was trained.
In this section, we describe a basic theory for estimating uncertainty due to model inadequacy by \emph{inflating} the uncertainty in the model's parameters.
The basic idea is to introduce fluctuations in the model's parameters with a scale comparable to the accuracy of the model.

In this formulation, we assume a collection of data $\{y_m\}_{m=1}^M$ and a parameterized family of model predictions $\{f_m(\theta)\}_{m=1}^M$.
In our notation, $\theta$ are the parameters of the model and $M$ denotes the number of observed data.
The data correspond to predictions that the model can make, which include primitive quantities such as energy, forces and stress, or more complex material properties such as equilibrium lattice constants or thermal conductivity.
Data for these quantities can be obtained experimentally or from more accurate first-principles calculations, such as density functional theory (DFT).
A commonly used IP fitting method, force-matching, uses the energy, forces, and stress for a set of atomic configurations with DFT calculations as data\cite{Ercolessi_Adams_1994, Fellinger_Park_Wilkins_2010, Wen_Shirodkar_Plechas_Kaxiras_Elliott_Tadmor_2017}.
However, data for other material properties, such as lattice parameters\cite{Ercolessi_Adams_1994} and thermal conductivity\cite{Vohra_Nobakht_Shin_Mahadevan_2018, Vohra_Mahadevan_2019}, are also used.

To compare model predictions against the data, we introduce a \emph{cost} (or loss) function.
The most commonly used cost is (weighted) least squares,
\begin{equation}
    \label{eq:cost}
    C(\theta) = \frac{1}{2} \sum_{m=1}^M w_m (y_m - f_m(\theta))^2,
\end{equation}
where $w_m$ are the weights.

Selection of the weights is an important first step in quantifying uncertainty.
Since potentials are often trained for material properties that carry different physical units, such as eV for energy and eV/$\angstrom$ for forces, data should be appropriately weighted to put them on a common scale.
Functionally, the role of the weights is to quantify the relative target accuracy for each of the model's predictions.
When random errors in the data are the dominant source of uncertainty, $w_m$ are often taken to be the inverse square of the standard error (error bars) of the experiments.
However, since the dominant source of error in molecular modeling is the functional form of the IP, rather than errors in the data, some expert judgment needs to be used.
Lenosky et al.\ \cite{Lenosky_Kress_Kwon_Voter_Edwards_Richards_Yang_Adams_1997} suggest setting the weights to a fraction of the values of the data with a padding term to deal with near-zero values, so that each data point has an unique weight.
In this case, the weights are computed (with some notational change) as
\begin{equation}
    \label{eq:weights}
    w_m^{-1} = c_1^2 + c_2^2 \Vert y_m \Vert^2,
\end{equation}
where $c_1$ and $c_2$ are the padding term and fractional scale of the data, respectively.
The inverse weight in Eq.~\eqref{eq:weights} should be understood as a permissible tolerance on each data point, with $c_1$ and $c_2$ acting as absolute and relative tolerance, respectively.
In our approach, only the relative values of the weights matter, as we later scale the weights uniformly to estimate the magnitude of the model inadequacy.

The best fit parameters, $\hat{\theta}$, are those that minimize the cost function:
\begin{equation}
  \label{eq:bestfit}
  \hat{\theta} = \arg \min_\theta C(\theta).
\end{equation}
There are many optimization algorithms that can be used to estimate the best fit.
For machine learning potentials, stochastic gradient descent is typically used\cite{Behler_2016}, while the Levenberg--Marquardt algorithm is particularly effective for training empirical potentials\cite{Transtrum_Sethna_2012,Wen_Li_Brommer_Elliott_Sethna_Tadmor_2016}, especially when the potential is sloppy (as seems to generally be the case)\cite{whynonlinearfits}.

Having found the best fit, we quantify uncertainty in these estimated parameter values by considering sub-optimal parameter values within some tolerance.
We, therefore, give a statistical interpretation to the optimization problem.
The cost function is the negative log-likelihood\footnote{The likelihood function describes the probability of obtaining the observed data for a given set of model parameters, $P\left(\mathbf{y} \middle| \theta\right)$.} of the model parameters given the data:
\begin{equation}
    \label{eq:likelihood}
    L\left(\theta \middle| \mathbf{y}\right) \propto \exp(-C(\theta)).
\end{equation}
For weighted least squares, the likelihood in Eq.~\eqref{eq:likelihood} corresponds to the assumption that the data are generated by the model with some additional, random noise:
\begin{equation}
    \label{eq:data_model_noise}
    y_m = f_m(\theta^*) + \epsilon_m,
\end{equation}
where $f_m$ is the $m^{\text{th}}$ model prediction, $\theta^*$ are the ``true'' parameter values, and $\epsilon_m$ is an error term modeled as a Gaussian random variable with zero mean and variance $\sigma_m^2 = 1/w_m$\cite{Christensen_Bligaard_Jacobsen_2020}.
Note that the best fit parameters $\hat{\theta}$ are an estimate of $\theta^*$.

In molecular modeling, the dominant source of error originates from the IP's functional form, i.e., it contains errors due to its limited scope and missing physics.
Thus, we decompose $\epsilon_m$ in Eq.~(\ref{eq:data_model_noise}) as a combination of model inadequacy\cite{Kennedy_OHagan_2001}, $b_m$, and errors in the data (e.g., inaccuracies in the DFT values), $\xi_m$:
\begin{equation}
    \label{eq:data_model_bias_noise}
    y_m = f_m(\theta^*) + b_m + \xi_m.
\end{equation}
Considerable recent effort has been exerted to rigorously estimate the errors associated with DFT values\cite{Mortensen_Kaasbjerg_Frederiksen_Norskov_Sethna_Jacobsen_2005, Christensen_Bligaard_Jacobsen_2020}, corresponding to scale of $\xi_m$.
However, in most molecular modeling applications, the bias $b_m$ is the dominant source of error, and a major focus of this paper.
In general, modeling the bias is an important, unsolved problem in UQ.

There are several suggestion on how to handle model bias, such as by directly improving the model or by applying statistical correction to the model prediction\cite{parameterinflation}; the latter is the focus of this paper.
The statistical correction is added by inflating the likelihood\cite{Frederiksen_Jacobsen_Brown_Sethna_2004, Petzold_Bligaard_Jacobsen_2012, Christensen_Bligaard_Jacobsen_2020}, modifying Eq.~\eqref{eq:likelihood} as
\begin{equation}
    \label{eq:likelihood_T}
    L(\theta | \mathbf{y}) \propto \exp(-C(\theta) / T).
\end{equation}
Here, $T > 1$ is a hyper-parameter that we adjust to account for inadequacies of the model.
Functionally,  this temperature is equivalent to uniformly scaling the weights in Eq.~\eqref{eq:cost}.
Note that the choice of $T$ does not affect the best fit parameter values, $\hat{\theta}$, but it will affect the uncertainty associated with those values.

To estimate the statistical uncertainty in the parameters corresponding to the likelihood in Eq.~\eqref{eq:likelihood_T}, we use a Bayesian approach known as Markov chain Monte Carlo (MCMC).
In the Bayesian framework, the parametric uncertainty is encoded in a \emph{posterior} probability distribution of parameters given data, $P\left(\theta \middle| \mathbf{y} \right)$.
The posterior is related to the likelihood by Bayes' theorem,
\begin{equation}
    P\left(\theta \middle| \mathbf{y}\right) \propto L\left(\theta \middle| \mathbf{y}\right) \times \pi(\theta),
    \label{eq:bayes}
\end{equation}
where $L(\theta \vert \mathbf{y})$ is the tempered likelihood in Eq.~\eqref{eq:likelihood_T} and $\pi(\theta)$ is the \emph{prior} distribution of the parameters.

The prior, $\pi(\theta)$, must be provided by the modeler and is another important input into the UQ formalism.
Nominally, it encodes the modeler's prior expectation for the values of the parameters in the model.
Common choices include uniform\cite{Angelikopoulos_Papadimitriou_Koumoutsakos_2012, Rizzi_Jones_Debusschere_Knio_2013, Messerly_Shirts_Kazakov_2018, Vohra_Nobakht_Shin_Mahadevan_2018, Vohra_Mahadevan_2019, Dhaliwal_Nair_Singh_2019a, Dhaliwal_Nair_Singh_2019b}, normal\cite{De_Simon_Iglesias_Jones_Wood_2018}, Jeffreys prior\cite{Rizzi_Najm_Debusschere_Sargsyan_Salloum_Adalsteinsson_Knio_2012}, and maximum entropy\cite{Cools-Ceuppens_Verstraelen_2017}.
However, there is rarely an obviously ``correct'' prior to choose.
At best, modelers may have a vague notion of a typical expected value or range for a parameter, while in other cases there may be no prior information at all.
Because of this ambiguity, we recommend that calculations be done for several choices of prior distributions to ensure that any conclusions are robust to this arbitrary choice.

% Analogy to stat mech
The remaining undefined quantity in Eq.~\eqref{eq:bayes} is the hyperparameter $T$.
To choose a reasonable sampling temperature, we invoke a formal analogy between Bayesian statistics and the Boltzmann distribution in statistical mechanics.
Writing the posterior as
\begin{equation}
    \label{eq:posterior_boltzmann}
    P\left(\theta \middle| \mathbf{y}\right) \propto \exp{\left( -(C(\theta) - T S(\theta)) / T \right)}
\end{equation}
suggests that cost is analogous to the energy while the (log) prior is analogous to the entropy, $S(\theta) = \log(\pi(\theta))$.
This analogy motivates a natural way to select the temperature in Eq.~\eqref{eq:likelihood_T} as an estimate of the scale of model bias.
We adjust the temperature to make the fluctuations of cost in the posterior distribution comparable to the best fit cost.
According to the equipartition theorem, each parameter mode in a harmonic model will contribute $T/2$, so Frederiksen et al.\ \cite{Frederiksen_Jacobsen_Brown_Sethna_2004, Petzold_Bligaard_Jacobsen_2012} advocate using
\begin{equation}
    \label{eq:T0}
    T_0 = 2C_0 / N,
\end{equation}
where $C_0 = C(\hat{\theta})$ is the cost at the best fit and $N$ is the number of parameters in the model.
The value of $C_0$ is our best available estimate of the scale of the model's inadequacy.
This choice of temperature uniformly scales the weights in Eq.~\eqref{eq:cost}, inflating the error bars in the cost to be comparable in size to the minimal cost.

IPs are not harmonic models, so the specific choice given in Eq.\ref{eq:T0} is only a rough guideline\cite{parameterinflation}.
Recent studies have explored anharmonic effects in sloppy IPs\cite{parameterinflation, kurniawan2021bayesian}.
At high sampling temperatures, the posterior is dominated by entropy and becomes sensitive to the choice of prior.
The entropic contribution from the sloppy, degenerate modes can overwhelm the posterior and give biased predictions, as we demonstrate below for a specific example.
Because of these anharmonic effects, practitioners should consider ensembles for many temperatures and different choices of prior to explore the sensitivity of their conclusions to these arbitrary choices.

% Parallel tempering
Having defined all terms on the right-hand side of Eq.~(\ref{eq:bayes}), the posterior $P\left(\theta \middle| \mathbf{y}\right)$ can be sampled via some MCMC-based algorithm.
To efficiently sample at several temperatures, we invoke parallel-tempered MCMC (PTMCMC) methods.
Tempering the likelihood as in Eq.~\eqref{eq:likelihood_T} is commonly used in PTMCMC methods to improve the convergence rate of the sampling\cite{Earl_Deem_2005, Miller_Dunson_2019}.
These algorithms generate multiple Markov chains, each at different temperatures, and mix them with an appropriate probability to ensure convergence to the target posterior\cite{Vousden_Farr_Mandel_2016}.
Here, we use PTMCMC methods as part of our UQ framework to empirically assess the effects of sampling temperature.
In practice, we consider a chain of temperatures from $T=1$ up to a few times larger than $T_0$ (defined in Eq.~(\ref{eq:T0})).
These temperatures explore the transition from sampling at the target accuracy (set by $w_m$) to a more realistic estimate of the systematic error, accounting for model inadequacy by inflating the likelihood with $T_0$.

% Convergence
After a sufficient number of iterations, the distribution of MCMC samples will converge to the posterior\cite{Gelman_Rubin_1992}.
The multivariate potential scale reduction factor (PSRF), denoted by $\hat{R}^p$, is a common metric to assess convergence in MCMC chains.
The $\hat{R}^p$ compares the variance between and within independent chains by
\begin{equation}
    \label{eq:psrf}
    \hat{R}^p = \frac{K-1}{K} + \frac{J+1}{J} \lambda_\text{max} (W^{-1} B/K),
\end{equation}
where $J$ and $K$ are the numbers of chains and iterations, respectively, and $\lambda_\text{max}(A)$ is the largest eigenvalue of the matrix $A$.
The parameters, $B/K$ and $W$, are the variance between and within the independent chains, $\psi_j$, which are calculated by
\begin{equation}
    \label{eq:rhat_variances}
    \begin{aligned}
	\frac{B}{K} &= \frac{1}{J-1} \sum_{j=1}^J \left( \bar{\psi}_j - \bar{\psi} \right) \left( \bar{\psi}_j - \bar{\psi} \right)^T, \\
	W &= \frac{1}{J(K - 1)} \sum_{j=1}^J \sum_{k=1}^K \left( \psi_{jk} - \bar{\psi}_j \right) \left( \psi_{jk} - \bar{\psi}_j \right)^T,	
    \end{aligned}
\end{equation}
with $\psi_{jk}$ denoting the $k^{\text{th}}$ iteration of the $j^{\text{th}}$ chain, $\bar{\psi}_j$ denoting the average of the $j^{\text{th}}$ chain, and $\bar{\psi}$ denoting the average over all chains and iterations.
The value of $\hat{R}^p$ monotonically decreases to one as $K \to \infty$ and the chains converge to the stationary distribution.
In practice, a common threshold is in the range of 1.05 to 1.1\cite{Gelman_Rubin_1992, Brooks_Gelman_1998}, although higher thresholds have also been used\cite{vats2021revisiting}.

\subsection{The OpenKIM project}
\label{subsec:openkim}

% OpenKIM and KIM API
The Open Knowledgebase of Interatomic Models (OpenKIM) is a National Science Foundation (NSF)-funded cyberinfrastructure project that aims to create an organized framework for the application of IPs that yields publicly accessible and reproducible results\cite{Tadmor_Elliott_Sethna_Miller_Becker_2011}.
In OpenKIM terms, a ``model'' refers to a standardized computer implementation of an IP with a fixed set of parameter values.
OpenKIM models are publicly available in an open-source online repository at \href{https://openkim.org/}{https://openkim.org/} \cite{kim_items}.
IPs archived in OpenKIM conform to the KIM application programming interface (API)\cite{elliott:tadmor:2011}, which allows them to work seamlessly with multiple molecular simulation codes \cite{kim_codes}.

To facilitate the development of new IPs, the OpenKIM project has developed the KIM-based Learning-Integrated Fitting Framework (KLIFF) \cite{Wen_Afshar_Elliott_Tadmor_2022,kliff-git}.
KLIFF is a general purpose fitting framework written in Python for both physics-based and machine learning IPs.
By default, KLIFF employs a force-matching algorithm to train an IP that uses the energy, forces, and stresses (if available) for a set of atomic configurations.
The inclusion of other material properties in the cost function is also possible. 
Several built-in optimizers are provided (such as the Levenberg--Marquardt algorithm), as well as many others available through the SciPy package\cite{2020SciPy-NMeth}.
IPs trained with KLIFF conform to the KIM API and therefore are automatically compatible with multiple molecular simulation codes as noted above.

% ColabFit
Finally, KLIFF is integrated with the ColabFit project\cite{colabfit} providing it with access to a large repository of vetted, high-quality training data for IP fitting.

\section{UQ Extensions to KLIFF}
\label{sec:solution}

\begin{figure}
    \centering
    \includegraphics[width=0.5\textwidth]{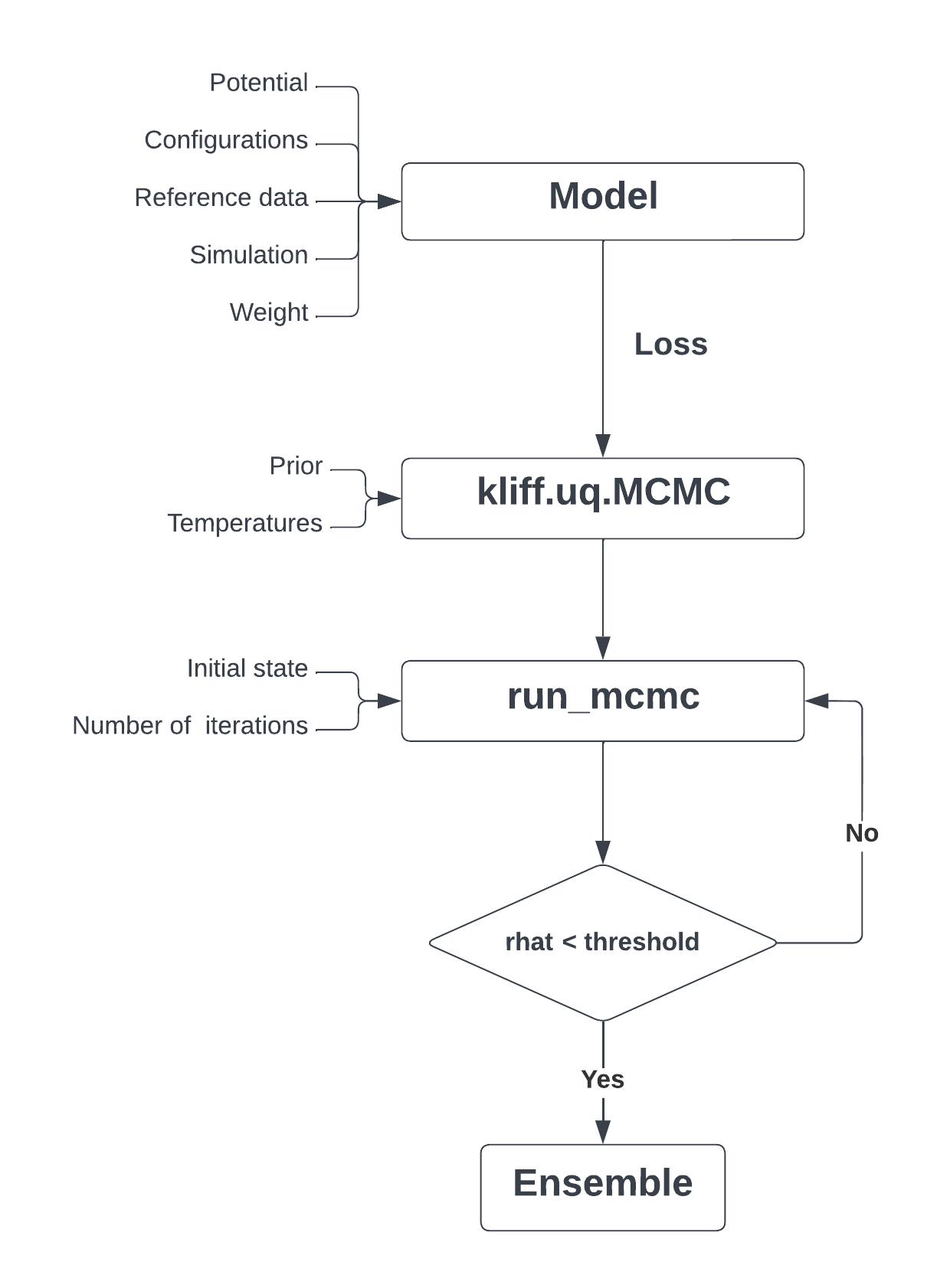}
    \caption{
	UQ framework implemented in KLIFF.
	The UQ process starts with the construction of the model and the loss function, followed by the creation of a posterior sampler with the \texttt{kliff.uq.MCMC} class.
	Following that, MCMC sampling is performed with the \texttt{run\_mcmc} function, terminating once the samples converge to a stationary distribution, e.g. when $\hat{R}^p$ is below an acceptable threshold.
	The distribution of the parameters is deduced from the samples.
    }
    \label{fig:workflow}
\end{figure}

The KLIFF package provides a convenient environment for fitting IPs.
In this work, we introduce a new toolkit for use with KLIFF that provides a framework that facilitates transparent, reproducible UQ analysis of IPs.
As MCMC is the UQ method of choice in molecular modeling, we focus on this method first with the intent to add alternative sampling methods in the future.
A typical workflow is as follows:
First, a modeler selects (or defines) the model and formulates the posterior sampler by instantiating \texttt{kliff.uq.MCMC}, which requires the specification of a prior and sampling temperature.
Next, an MCMC simulation is performed with the \texttt{run\_mcmc} function to generate ensembles of parameters drawn from the target Bayesian posterior distribution.
This process is represented in Fig.~\ref{fig:workflow} and further discussed below.

% Defining the model
The first step in this workflow is to create the model.
The model comprises the parameterized IP as well as atomic configurations, reference data to calibrate the IP parameters, and a simulation that calculates the corresponding material properties.
KLIFF is used to construct a family of parameterized models based on an OpenKIM IP and read the configuration files that contain the atomic configurations and reference data to calibrate the IP.
By default, KLIFF uses a simulation that computes the energy, forces, and stress for each configuration, although extending KLIFF to use other simulation and the corresponding material properties is also possible (see Ref.~\cite{Wen_Afshar_Elliott_Tadmor_2022}).
Then, KLIFF uses this information to construct a weighted least squares cost function for the model (see Eq.~\eqref{eq:cost}).
An example of this process is shown in the Python script in Fig.~\ref{fig:model}.

\begin{figure}[!ht]
    \hrulefill
    % \inputminted[fontsize=\footnotesize]{python}{listings/listing1.txt}
\begin{Verbatim}[commandchars=\\\{\}, fontsize=\footnotesize]
\PYG{k+kn}{import} \PYG{n+nn}{os}
\PYG{k+kn}{import} \PYG{n+nn}{numpy} \PYG{k}{as} \PYG{n+nn}{np}

\PYG{k+kn}{from} \PYG{n+nn}{kliff.calculators} \PYG{k+kn}{import} \PYG{n}{Calculator}
\PYG{k+kn}{from} \PYG{n+nn}{kliff.dataset} \PYG{k+kn}{import} \PYG{n}{Dataset}
\PYG{k+kn}{from} \PYG{n+nn}{kliff.dataset.weight} \PYG{k+kn}{import} \PYG{p}{(}
    \PYG{n}{MagnitudeInverseWeight}\PYG{p}{,}
\PYG{p}{)}
\PYG{k+kn}{from} \PYG{n+nn}{kliff.loss} \PYG{k+kn}{import} \PYG{n}{Loss}
\PYG{k+kn}{from} \PYG{n+nn}{kliff.models} \PYG{k+kn}{import} \PYG{n}{KIMModel}
\PYG{k+kn}{from} \PYG{n+nn}{kliff.models.parameter\PYGZus{}transform} \PYG{k+kn}{import} \PYG{p}{(}
    \PYG{n}{LogParameterTransform}\PYG{p}{,}
\PYG{p}{)}
\PYG{k+kn}{from} \PYG{n+nn}{kliff.utils} \PYG{k+kn}{import} \PYG{n}{download\PYGZus{}dataset}

\PYG{c+c1}{\PYGZsh{} Instantiate a transformation class to do the log}
\PYG{c+c1}{\PYGZsh{} parameter transform}
\PYG{n}{param\PYGZus{}names} \PYG{o}{=} \PYG{p}{[}\PYG{l+s+s2}{\PYGZdq{}A\PYGZdq{}}\PYG{p}{,} \PYG{l+s+s2}{\PYGZdq{}B\PYGZdq{}}\PYG{p}{,} \PYG{l+s+s2}{\PYGZdq{}sigma\PYGZdq{}}\PYG{p}{,} \PYG{l+s+s2}{\PYGZdq{}lambda\PYGZdq{}}\PYG{p}{,} \PYG{l+s+s2}{\PYGZdq{}gamma\PYGZdq{}}\PYG{p}{]}
\PYG{n}{params\PYGZus{}transform} \PYG{o}{=} \PYG{n}{LogParameterTransform}\PYG{p}{(}
    \PYG{n}{param\PYGZus{}names}
\PYG{p}{)}

\PYG{c+c1}{\PYGZsh{} Instantiate the model and set the potential}
\PYG{n}{model} \PYG{o}{=} \PYG{n}{KIMModel}\PYG{p}{(}
    \PYG{l+s+s2}{\PYGZdq{}SW\PYGZus{}StillingerWeber\PYGZus{}1985\PYGZus{}Si\PYGZus{}\PYGZus{}MO\PYGZus{}405512056662\PYGZus{}006\PYGZdq{}}\PYG{p}{,}
    \PYG{n}{params\PYGZus{}transform}\PYG{p}{,}
\PYG{p}{)}

\PYG{c+c1}{\PYGZsh{} Set the tunable parameters and the initial guess}
\PYG{n}{opt\PYGZus{}params} \PYG{o}{=} \PYG{p}{\PYGZob{}}
    \PYG{n}{name}\PYG{p}{:} \PYG{p}{[[}\PYG{l+s+s2}{\PYGZdq{}default\PYGZdq{}}\PYG{p}{]]} \PYG{k}{for} \PYG{n}{name} \PYG{o+ow}{in} \PYG{n}{param\PYGZus{}names}
\PYG{p}{\PYGZcb{}}
\PYG{n}{model}\PYG{o}{.}\PYG{n}{set\PYGZus{}opt\PYGZus{}params}\PYG{p}{(}\PYG{o}{**}\PYG{n}{opt\PYGZus{}params}\PYG{p}{)}

\PYG{c+c1}{\PYGZsh{} Get the dataset and set the weights}
\PYG{n}{dataset\PYGZus{}path} \PYG{o}{=} \PYG{n}{download\PYGZus{}dataset}\PYG{p}{(}\PYG{l+s+s2}{\PYGZdq{}Si\PYGZus{}training\PYGZus{}set\PYGZdq{}}\PYG{p}{)}
\PYG{n}{dataset\PYGZus{}path} \PYG{o}{/=} \PYG{l+s+s2}{\PYGZdq{}varying\PYGZus{}alat\PYGZdq{}}
\PYG{c+c1}{\PYGZsh{} Instantiate the weight class}
\PYG{n}{weight} \PYG{o}{=} \PYG{n}{MagnitudeInverseWeight}\PYG{p}{(}
    \PYG{c+c1}{\PYGZsh{} Each key in weight\PYGZus{}params contains a list}
    \PYG{c+c1}{\PYGZsh{} [c\PYGZus{}1, c\PYGZus{}2]}
    \PYG{n}{weight\PYGZus{}params}\PYG{o}{=}\PYG{p}{\PYGZob{}}
        \PYG{l+s+s2}{\PYGZdq{}energy\PYGZus{}weight\PYGZus{}params\PYGZdq{}}\PYG{p}{:} \PYG{p}{[}\PYG{l+m+mf}{0.0}\PYG{p}{,} \PYG{l+m+mf}{0.1}\PYG{p}{],}
        \PYG{l+s+s2}{\PYGZdq{}forces\PYGZus{}weight\PYGZus{}params\PYGZdq{}}\PYG{p}{:} \PYG{p}{[}\PYG{l+m+mf}{1e\PYGZhy{}2}\PYG{p}{,} \PYG{l+m+mf}{0.1}\PYG{p}{],}
    \PYG{p}{\PYGZcb{}}
\PYG{p}{)}
\PYG{c+c1}{\PYGZsh{} Read the configurations and reference data from}
\PYG{c+c1}{\PYGZsh{} the dataset}
\PYG{n}{tset} \PYG{o}{=} \PYG{n}{Dataset}\PYG{p}{(}\PYG{n}{dataset\PYGZus{}path}\PYG{p}{,} \PYG{n}{weight}\PYG{o}{=}\PYG{n}{weight}\PYG{p}{)}
\PYG{n}{configs} \PYG{o}{=} \PYG{n}{tset}\PYG{o}{.}\PYG{n}{get\PYGZus{}configs}\PYG{p}{()}

\PYG{c+c1}{\PYGZsh{} Create a calculator, which consists of simulations}
\PYG{c+c1}{\PYGZsh{} to compute material properties}
\PYG{n}{calc} \PYG{o}{=} \PYG{n}{Calculator}\PYG{p}{(}\PYG{n}{model}\PYG{p}{)}
\PYG{c+c1}{\PYGZsh{} Set the configurations to use by the calculator}
\PYG{n}{ca} \PYG{o}{=} \PYG{n}{calc}\PYG{o}{.}\PYG{n}{create}\PYG{p}{(}\PYG{n}{configs}\PYG{p}{)}

\PYG{c+c1}{\PYGZsh{} Instantiate the loss function}
\PYG{n}{residual\PYGZus{}data} \PYG{o}{=} \PYG{p}{\PYGZob{}}\PYG{l+s+s2}{\PYGZdq{}normalize\PYGZus{}by\PYGZus{}natoms\PYGZdq{}}\PYG{p}{:} \PYG{k+kc}{False}\PYG{p}{\PYGZcb{}}
\PYG{n}{loss} \PYG{o}{=} \PYG{n}{Loss}\PYG{p}{(}\PYG{n}{calc}\PYG{p}{,} \PYG{n}{residual\PYGZus{}data}\PYG{o}{=}\PYG{n}{residual\PYGZus{}data}\PYG{p}{)}

\PYG{c+c1}{\PYGZsh{} Train the model}
\PYG{n}{loss}\PYG{o}{.}\PYG{n}{minimize}\PYG{p}{(}\PYG{n}{method}\PYG{o}{=}\PYG{l+s+s2}{\PYGZdq{}lm\PYGZdq{}}\PYG{p}{)}
\end{Verbatim}
    \hrulefill
    \caption{Example Python script for constructing a model using KLIFF.}
    \label{fig:model}
\end{figure}

% Weight
In previous versions (version 0.3.3 or earlier), KLIFF allowed for a single weight for each type of material property in the cost function, e.g., a single weight for all energies, a single weight for all forces, and so on.
To enable more flexible UQ sampling, we expand KLIFF to enable specification of custom weights for each data point.
Specific support is added for the weight calculation as defined in Eq.~\eqref{eq:weights} that takes arguments $c_1$ and $c_2$.
The default values are $c_1=1.0$ and $c_2=0.0$, which corresponds to a uniform weight for each data point.
(See \verb|MagnitudeInverseweight| in Fig.~\ref{fig:model} for setting $c_1$ and $c_2$ to values other than the defaults.)
Alternatively, users can define their own method to compute the weights.
This update is included in KLIFF version 0.4.0.

% Our implementation and ptemcee
The UQ framework is implemented as module \texttt{uq} in KLIFF.
The posterior sampler is constructed by creating a \texttt{kliff.uq.MCMC} instance.
Internally, this action computes $T_0$, generates a temperature ladder, and defines the untempered log-likelihood and log-prior functions.
As a default, this class inherits from the sampler in the \texttt{ptemcee} Python package to perform PTMCMC.
The algorithm simulates multiple chains (several at each sampling temperature) in parallel and mixes chains from different sampling temperatures to allow the MCMC walkers to explore a wider range of parameters.
The number of chains or walkers can be specified through an optional argument \texttt{nwalkers}; the default value is twice the number of parameters in the model.
This process is illustrated in the listing in Fig.~\ref{fig:uq_mcmc}.

\begin{figure}[!ht]
    \hrulefill
    % \inputminted[fontsize=\footnotesize]{python}{listings/listing2.txt}
\begin{Verbatim}[commandchars=\\\{\}, fontsize=\footnotesize]
\PYG{k+kn}{from} \PYG{n+nn}{kliff.uq} \PYG{k+kn}{import} \PYG{n}{MCMC}\PYG{p}{,} \PYG{n}{get\PYGZus{}T0}
\PYG{k+kn}{from} \PYG{n+nn}{multiprocessing} \PYG{k+kn}{import} \PYG{n}{Pool}

\PYG{n}{np}\PYG{o}{.}\PYG{n}{random}\PYG{o}{.}\PYG{n}{seed}\PYG{p}{(}\PYG{l+m+mi}{2022}\PYG{p}{)}

\PYG{c+c1}{\PYGZsh{} Get the dimensionality of the problem}
\PYG{c+c1}{\PYGZsh{} Number of parameters}
\PYG{n}{ndim} \PYG{o}{=} \PYG{n}{calc}\PYG{o}{.}\PYG{n}{get\PYGZus{}num\PYGZus{}opt\PYGZus{}params}\PYG{p}{()}
\PYG{n}{nwalkers} \PYG{o}{=} \PYG{l+m+mi}{2} \PYG{o}{*} \PYG{n}{ndim}  \PYG{c+c1}{\PYGZsh{} Number of parallel walkers}

\PYG{c+c1}{\PYGZsh{} Generate a temperature ladder}
\PYG{n}{T0} \PYG{o}{=} \PYG{n}{get\PYGZus{}T0}\PYG{p}{(}\PYG{n}{loss}\PYG{p}{)}
\PYG{n}{Tladder} \PYG{o}{=} \PYG{n}{np}\PYG{o}{.}\PYG{n}{sort}\PYG{p}{(}
    \PYG{n}{np}\PYG{o}{.}\PYG{n}{append}\PYG{p}{(}\PYG{n}{np}\PYG{o}{.}\PYG{n}{logspace}\PYG{p}{(}\PYG{l+m+mi}{0}\PYG{p}{,} \PYG{l+m+mi}{7}\PYG{p}{,} \PYG{l+m+mi}{15}\PYG{p}{),} \PYG{n}{T0}\PYG{p}{)}
\PYG{p}{)}
\PYG{n}{ntemps} \PYG{o}{=} \PYG{n+nb}{len}\PYG{p}{(}\PYG{n}{Tladder}\PYG{p}{)}  \PYG{c+c1}{\PYGZsh{} Number of temperatures}

\PYG{c+c1}{\PYGZsh{} Instantiate a sampler}
\PYG{n}{sampler} \PYG{o}{=} \PYG{n}{MCMC}\PYG{p}{(}
    \PYG{n}{loss}\PYG{p}{,}
    \PYG{n}{nwalkers}\PYG{o}{=}\PYG{n}{nwalkers}\PYG{p}{,}
    \PYG{n}{logprior\PYGZus{}args}\PYG{o}{=}\PYG{p}{(}\PYG{n}{np}\PYG{o}{.}\PYG{n}{tile}\PYG{p}{([}\PYG{o}{\PYGZhy{}}\PYG{l+m+mi}{8}\PYG{p}{,} \PYG{l+m+mi}{8}\PYG{p}{],} \PYG{p}{(}\PYG{n}{ndim}\PYG{p}{,} \PYG{l+m+mi}{1}\PYG{p}{)),),}
    \PYG{n}{Tladder}\PYG{o}{=}\PYG{n}{Tladder}\PYG{p}{,}
    \PYG{c+c1}{\PYGZsh{} Other keyword arguments for ptemcess.Sampler}
    \PYG{n}{random}\PYG{o}{=}\PYG{n}{np}\PYG{o}{.}\PYG{n}{random}\PYG{o}{.}\PYG{n}{RandomState}\PYG{p}{(}\PYG{l+m+mi}{2022}\PYG{p}{),}
\PYG{p}{)}
\PYG{c+c1}{\PYGZsh{} Declare multiprocessing pool for parallel}
\PYG{c+c1}{\PYGZsh{} computing}
\PYG{n}{sampler}\PYG{o}{.}\PYG{n}{pool} \PYG{o}{=} \PYG{n}{Pool}\PYG{p}{(}\PYG{n}{processes}\PYG{o}{=}\PYG{n}{nwalkers}\PYG{p}{)}
\end{Verbatim}
    \hrulefill
    \caption{
        Example Python script for constructing the sampler with the \texttt{kliff.uq.MCMC} class.
        See Fig.~\ref{fig:model} for an example of how to define the calculator and loss function.
    }
    \label{fig:uq_mcmc}
\end{figure}

% Specifying the likelihood and prior
The arguments to instantiate \texttt{kliff.uq.MCMC} are (1) a \texttt{kliff.loss.Loss} instance, which defines the \emph{untempered} ($T=1$) likelihood function in Eq.~\eqref{eq:likelihood}, (2) the prior, and (3) the sampling temperatures.
Since uniform priors are a common choice, the constructor implements this by default.
In this case, the boundaries of the prior support need to be specified by the user via the \texttt{logprior\_args} argument.
However, the user can also pass a custom prior as the \texttt{logprior\_fn} argument.

% Sampling temperatures
There are two options to specify the temperature ladder.
First, the user may specify the number of temperatures (\texttt{ntemps}) and the ratio between the maximum temperature and $T_0$ in Eq.~\eqref{eq:T0} (\texttt{Tmax\_ratio}).
In this case, an internal function generates a logarithmically spaced temperature between $T=1.0$ and $T=T_{\text{max\_ratio}} \times T_0$, inclusive.
Alternatively, the user may specify the complete list of temperature values (\texttt{Tladder}).
When \texttt{Tladder} is specified, then this list overrides the values passed for \texttt{ntemps} and \texttt{Tmax\_ratio}.

The UQ implementation in KLIFF supports parallelization over the configurations for the cost function evaluation and over the walkers for the MCMC sampling.
However, the current implementation only supports OpenMP-style parallelization for the cost function evaluation and both OpenMP and MPI for the MCMC sampling, with a future work to allow MPI in the earlier.
In this paper we only show an example of parallelization in the Bayesian sampling, which is done by declaring a \texttt{multiprocessing} pool after instantiating \texttt{kliff.uq.MCMC}.
The optimal number of parallel processes in this case is the product of the numbers of the sampling temperatures and the walkers.

% How to run the sampling
After constructing the posterior sampler, MCMC sampling is run by calling the \texttt{run\_mcmc} method of the \texttt{kliff.uq.MCMC} class (see Fig.~\ref{fig:run_mcmc}).
This function requires the initial position of each walker as an $L \times J \times N$ array, where $L, J,$ and $N$ are the number of temperatures, walkers, and parameters, respectively.
The user also needs to specify the number of iterations to run the MCMC simulation.

\begin{figure}[!ht]
    \hrulefill
    % \inputminted[fontsize=\footnotesize]{python}{listings/listing3.txt}
\begin{Verbatim}[commandchars=\\\{\}, fontsize=\footnotesize]
\PYG{c+c1}{\PYGZsh{} Initial starting points for each walker}
\PYG{n}{p0} \PYG{o}{=} \PYG{n}{np}\PYG{o}{.}\PYG{n}{random}\PYG{o}{.}\PYG{n}{uniform}\PYG{p}{(}
    \PYG{n}{low}\PYG{o}{=\PYGZhy{}}\PYG{l+m+mf}{6.0}\PYG{p}{,}
    \PYG{n}{high}\PYG{o}{=}\PYG{l+m+mf}{6.0}\PYG{p}{,}
    \PYG{n}{size}\PYG{o}{=}\PYG{p}{(}\PYG{n}{ntemps}\PYG{p}{,} \PYG{n}{nwalkers}\PYG{p}{,} \PYG{n}{ndim}\PYG{p}{),}
\PYG{p}{)}

\PYG{c+c1}{\PYGZsh{} Run MCMC}
\PYG{n}{sampler}\PYG{o}{.}\PYG{n}{run\PYGZus{}mcmc}\PYG{p}{(}\PYG{n}{p0}\PYG{p}{,} \PYG{l+m+mi}{150000}\PYG{p}{)}
\PYG{n}{sampler}\PYG{o}{.}\PYG{n}{pool}\PYG{o}{.}\PYG{n}{close}\PYG{p}{()}
\end{Verbatim}
    \hrulefill
    \caption{Example Python script for using the \texttt{run\_mcmc} function to perform sampling.}
    \label{fig:run_mcmc}
\end{figure}

% Convergence
The convergence of the parameter chains is assessed by calculating $\hat{R}^p$.
In our implementation, this is realized by the function \texttt{kliff.uq.rhat}, demonstrated by the listing in Fig.~\ref{fig:convergence}.
This function takes an array containing the MCMC samples for one sampling temperature.
The $\hat{R}^p$ is calculated for each sampling temperature separately.
Note that in practice, some sampling temperatures may converge much sooner than others.
If the resulting values are larger than some threshold (typically 1.1), then the MCMC algorithm should continue to iterate.
If $\hat{R}^p$ is less than the target threshold, the samples are assumed to have converged to the posterior and the calculation is terminated.

\begin{figure}[!ht]
    \hrulefill
    % \inputminted[fontsize=\footnotesize]{python}{listings/listing4.txt}
\begin{Verbatim}[commandchars=\\\{\}, fontsize=\footnotesize]
\PYG{k+kn}{from} \PYG{n+nn}{kliff.uq} \PYG{k+kn}{import} \PYG{n}{rhat}

\PYG{c+c1}{\PYGZsh{} Retrieve the samples}
\PYG{c+c1}{\PYGZsh{} Set the burn\PYGZhy{}in time and thinning factor}
\PYG{n}{burnin}\PYG{p}{,} \PYG{n}{thin} \PYG{o}{=} \PYG{l+m+mi}{10000}\PYG{p}{,} \PYG{l+m+mi}{200}
\PYG{n}{samples} \PYG{o}{=} \PYG{n}{sampler}\PYG{o}{.}\PYG{n}{chain}\PYG{p}{[:,} \PYG{p}{:,} \PYG{n}{burnin}\PYG{p}{::}\PYG{n}{thin}\PYG{p}{]}

\PYG{c+c1}{\PYGZsh{} Assess convergence by computing rhat}
\PYG{n}{rhat\PYGZus{}array} \PYG{o}{=} \PYG{n}{np}\PYG{o}{.}\PYG{n}{empty}\PYG{p}{(}\PYG{n}{ntemps}\PYG{p}{)}
\PYG{k}{for} \PYG{n}{tidx} \PYG{o+ow}{in} \PYG{n+nb}{range}\PYG{p}{(}\PYG{n}{ntemps}\PYG{p}{):}
    \PYG{n}{rhat\PYGZus{}array}\PYG{p}{[}\PYG{n}{tidx}\PYG{p}{]} \PYG{o}{=} \PYG{n}{rhat}\PYG{p}{(}\PYG{n}{samples}\PYG{p}{[}\PYG{n}{tidx}\PYG{p}{])}
\end{Verbatim}
    \hrulefill
    \caption{Example Python script for using \texttt{kliff.uq.rhat} function to compute $\hat{R}^p$ as a convergence assessment tool.}
    \label{fig:convergence}
\end{figure}

This UQ framework is integrated in KLIFF and we provide some examples in an online \href{https://gitlab.com/yonatank93/kliff_uq}{repository}\cite{gitlab_kliff_uq}.
In the next section, we describe the use and interpretation of a UQ calculation for a Stillinger--Weber potential.

\section{Results}
\label{sec:results}

% Describe the setup: model, data, weight, prior, temperatures (maybe)
% IP: SW potential
Having described the basic interface to this UQ toolkit, we now consider the concrete example given in the combined listings in Figs.~\ref{fig:model}--\ref{fig:convergence}.
In this section we discuss the specific model and MCMC setup in these scripts.
We then present and discuss the sampling results and some of the subtleties associated with their interpretation.

This example is based on the Stillinger--Weber potential in OpenKIM\cite{OpenKIM_SW_driver, OpenKIM_SW}. This is a cluster potential originally introduced to model silicon \cite{Stillinger-Weber}.
For a system with $n$ atoms, the potential energy of atom $i$, $\mathcal{V}_i$, is given by
\begin{equation}
    \label{eq:atom_energy}
    \mathcal{V}_i = \sum_{j>i}^n \phi_2(r_{ij}) + \sum_{j\neq i}^n \sum_{\substack{k>j \\ k\neq i}}^n \phi_3 (r_{ij}, r_{ik}, \beta_{jik}),
\end{equation}
where $\phi_m$ denotes the $m$-body interactions.
The Stillinger--Weber potential includes both two-body and three-body interactions.
The two-body term only depends on the distance between atom $i$ and $j$, denoted by $r_{ij}$,
\begin{equation}
    \label{eq:sw_2body}
    \phi_2(r_{ij}) = A \left[ B \left( \frac{\sigma}{r_{ij}} \right)^p - \left( \frac{\sigma}{r_{ij}} \right)^q \right] \times \exp\left( \frac{\sigma}{r_{ij}-r^{\text{cut}}} \right).
\end{equation}
The three-body term additionally depends on $\beta_{ijk}$, which is the bond angle between the $i-j$ and $i-k$ bonds,
\begin{equation}
    \label{eq:sw_3body}
    \begin{aligned}
	\phi_3 \left( r_{ij}, r_{ik}, \beta_{jik} \right) =&~
	\lambda \left( \cos \beta_{jik} - \cos \beta^0 \right)^2 \\
	&\times \exp \left( \frac{\gamma}{r_{ij} - r^{\text{cut}}}
	    + \frac{\gamma}{r_{ik} - r^{\text{cut}}} \right).
    \end{aligned}
\end{equation}
This IP includes nine parameters: $A$, $B$, $\sigma$, $p$, $q$, $r^\text{cut}$, $\lambda$, $\beta^0$ and $\gamma$.
The total energy of the system $\mathcal{V}$ is
\begin{equation}
    \label{eq:total_energy}
    \mathcal{V} = \sum_{i=1}^n \mathcal{V}_i.
\end{equation}
The atomic forces are calculated by taking the negative gradient of Eq.~\eqref{eq:atom_energy} with respect to the atomic positions.

% Parameterization
In this example, we choose the tunable parameters to be $A, B, \sigma, \lambda,$ and $\gamma$.
The other parameters are fixed to the default values given in OpenKIM\cite{OpenKIM_SW}.
Physically, the tunable parameters $A$ and $\lambda$ set energy scales in the potential, and $\sigma$ and $\gamma$ set length scales.
To be physically relevant, these parameters are constrained to be positive.
Parameter $B$ controls the relative scale of the repulsive part of the interaction and, thus, is also constrained to be positive.
To enforce this constraint, we use a log-transform, i.e., the tunable parameters are $\theta = ( \log(A), \log(B), \log(\sigma), \log(\lambda), \log(\gamma) )$.

% Configurations, material properties, reference data, weights
The training set consists of the energies and forces of silicon in the diamond cubic crystal structure.
We use 400 atomic configurations, including stretched and compressed cells with random perturbations.
The reference energy and force data were generated using the environment-dependent interatomic potential (EDIP) \cite{OpenKIM_EDIP_driver, OpenKIM_EDIP_model, EDIP}.\footnote{Since our objective is to explore UQ rather than develop an accurate model for silicon, we take the EDIP potential to be the ``exact'' ground truth. This greatly reduces the computational cost of generating the training set relative to DFT.}
Next, we compute the weights using Eq.~(\ref{eq:weights}), with $c_1 = 10^{-2}$ eV/$\angstrom$ for the forces and zero for the energies, and $c_2 = 10^{-1}$ for both.

% Likelihood and choice of temperatures
The training produces the following best fit parameters for the SW potential:
\begin{equation*}
    \begin{aligned}
	A &= 15.27922231~\text{eV} \quad &&\lambda = 45.47927476~\text{eV} \\
	B &= 0.6032372 \quad &&\gamma = 2.51306949~\angstrom  \\
	\sigma &= 2.09420085~\angstrom. &&	
    \end{aligned}
\end{equation*}
The natural temperature for this model is $T_0=1.324$.
Notice that the reference data were generated using another IP.
Thus, while there is some model inadequacy, it is relatively small compared to some other real-world examples, which explains the low $T_0$ value; in practice $T_0$ is typically orders of magnitude larger\cite{kurniawan2021bayesian}.
In our analysis, we extend the temperature ladder to a much higher temperature to mimic typical effects users may encounter in real-world applications.

% Prior
We use a uniform prior where $\pi(\theta)$ is constant if $-8 < \theta < 8$ and zero otherwise.
The support of this prior is chosen to be sufficiently wide to ensure that sampling is not artificially restricted to regions near the best fit.

% Sampling and convergence
Next, PTMCMC sampling is performed for 150,000 iterations.
From each walker, we discard the first 10,000 steps as the burn-in time.
We thin the remaining chains by keeping every 200-th step to ensure uncorrelated samples.
From the remaining samples, the maximum value of $\hat{R}^p$ is 1.046.

\begin{figure*}[!ht]
    \centering
    \includegraphics[width=\textwidth]{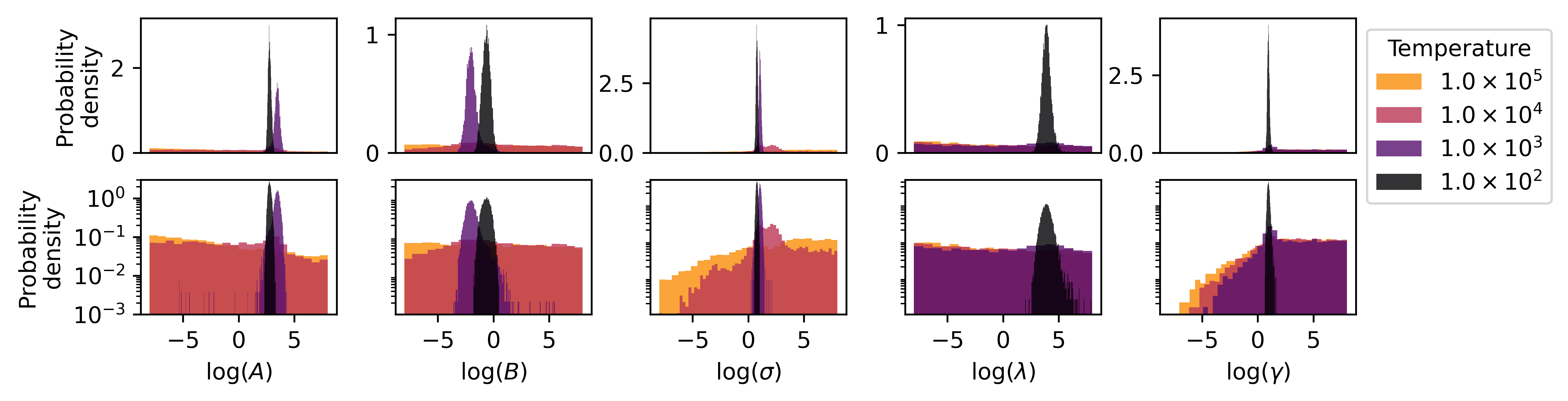}
    \caption{
	Marginal distributions of the MCMC samples (the projection of the joint distribution onto a single parameter axis) of the SW potential at several sampling temperatures, normalized by the number of samples.
	Each column shows the distribution of a each parameter, with the sampling temperatures shown by the different colors.
    On the second row, the distributions are presented in logarithmic scale on the vertical axis to bring out the details at higher sampling temperatures.
    }
    \label{fig:mcmc_results}
\end{figure*}

% Show the figures
The posterior distribution from which samples are drawn is a joint distribution for the five parameters.
Because we cannot directly visualize such a high dimensional space, it is common to instead plot marginal distributions.
The marginal distribution of a high-dimensional, joint probability distribution is the projection of the probability onto a lower dimensional subspace.\footnote{The marginal distribution of a parameter is calculated by summing the conditional distributions of that parameter over all possible values of the other parameters.}
We project the five-dimensional posterior probability distribution onto each of the one-dimensional parameter axes in Fig.~\ref{fig:mcmc_results}.
The marginal distributions at different sampling temperatures are shown in different colors and superimposed to enable direct comparison.
Each column corresponds to a different parameter in the potential.
In the second row, a log scale is used for the vertical axis to bring out the details of the sparser distributions that result at higher sampling temperatures.

% Brief discussion on the figures
Examining Fig.~\ref{fig:mcmc_results}, first note the general trend that distributions becomes wider as the temperature increases.
This is expected, since higher temperatures correspond to smaller weights, i.e., larger effective error bars, in Eq.~\eqref{eq:cost}.
However, the effect is parameter-dependent.  
For example, consider the distributions of $\log(\lambda)$ and $\log(\gamma)$ at $T = 10^2$.
At this temperature, the distributions are relatively localized around their best fit values.
However, at the next highest temperature, $T=10^3$, the distributions extend to the boundary of their respective priors.
This phenomenon, in which the marginal posterior distribution of some parameters abruptly transitions away from being localized at a specific sampling temperature, is called \emph{parameter evaporation}.
Inspecting Fig.~\ref{fig:mcmc_results} reveals that $\log(\lambda)$ and $\log(\gamma)$ evaporate around $T=10^3$ while parameters $A$ and $B$ evaporate around $T=10^4$.
At a sufficiently high sampling temperature, all parameters would evaporate, regardless of the prior, as show in Ref.~\cite{kurniawan2021bayesian} for SW potential for a molybdenum disulfide system.

Parameter evaporation has been observed in Ref.~\cite{kurniawan2021bayesian} for IPs and has important implications for UQ analysis.
To see this, first notice that while the distribution for $A$ and $B$ remain localized at $T=10^3$, they have shifted relative to their distributions at $T = 10^2$, i.e., they are localized around different values.
We can explain this shift in terms of the evaporation of the other parameters.
At $T = 10^3$, the range of values sampled for the parameters $\lambda$ and $\gamma$ are very different from those at $T = 10^2$.
Consequently, the sampled values for the parameters $A$ and $B$ shift to minimize the cost with this more diffuse distribution of $\lambda$ and $\gamma.$
That is to say, the values for parameters $A$ and $B$ are biased as a consequence of having the sub-optimal values of $\lambda$ and $\gamma$.
While this shift is not inherently problematic, notice that the distributions for $A$ and $B$ at $T = 10^3$ have very little overlap with their counterparts at $T = 10^2$.
Because the lower temperature distribution is dominated by samples near the best fit parameters, we infer that the higher temperature distribution has very few samples near the best fit.
This is potentially problematic since it implies that parameter values that best fit the data are not represented in the sample.

\begin{figure}[!ht]
    \centering
    \includegraphics[width=\columnwidth]{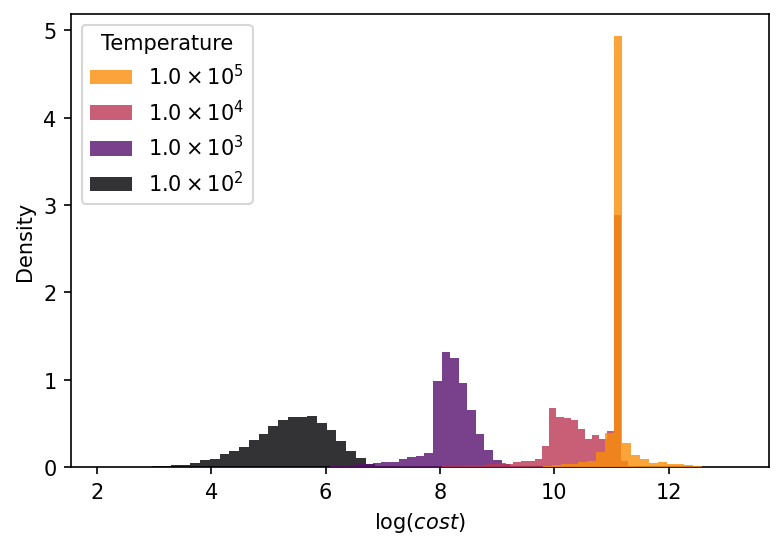}
    \caption{
	Distribution of cost at each sampling temperature.
    }
    \label{fig:mcmc_cost}
\end{figure}

Figure~\ref{fig:mcmc_cost} shows the distribution of the untempered costs at each sampling temperature.
As expected, the average value of the cost increases at higher temperatures.
However, this increase is not the result of stretching the distribution, as is the case for linear regression model.
Rather, the entire distribution shifts to the right at each rung in the temperature ladder.
As the temperature increases and parameters evaporate, the posterior is dominated by regions of the parameter space that are poor fits to the data.
These are regions of parameter space in which the prior places considerable weight, that is, they are high-entropy regions.
This is a nonlinear effect due to the subtle interplay between the sampling temperature, the prior, and the degenerate modes of sloppy models.
The resulting posterior distribution can depend very strongly on details of the problem, such as the choice of prior, the error bars for the data, and the sampling temperature.

In general, we recommend that practitioners check their results for robustness for several priors over a range of temperatures.
In this work, we have used uniform priors in log-transformed parameters as a pedagogical example.
This is a reasonable choice in general; however, we could have also used uniform priors in the original, untransformed parameterization as well as Gaussian priors in both log-transformed and untransformed parameters.
We caution against using Jeffreys prior, as its interaction with the degenerate modes can lead to strong biases\cite{quinn2021information}.
In this example, we have sampled up to a relatively high temperature relative to $T_0$ as defined in Eq.~\eqref{eq:T0}.
This was also a pedagogical choice to illustrate important effects and mimic more realistic examples (for example, previous work has seen values of $T_0 > 10^6$\cite{kurniawan2021bayesian}).  
In practice, we suggest including sampling temperatures up to a few times larger than $T_0$.
The highest sampling temperatures are not for the purposes of UQ, but including them in the sampling algorithm improves convergence rates.
For UQ analysis, we suggest considering ensembles generated by temperatures from $50\%$ below to $50\%$ above $T_0$.

IP developers should also use uncertainty quantification tools throughout the model development cycle.
For example, they could extend the training data to better constrain the sloppy parameters.
Alternatively, they could use model reduction methods\cite{MBAM} to remove the degenerate modes from the model.
For the latter, care must be taken to not remove unidentifiable parameters that would be important for downstream applications.

\section{Conclusion}
\label{sec:conclusion}

% What we did
In this work, we describe a UQ toolkit that extends the KLIFF package as part of the OpenKIM environment.
This toolkit provides a framework that facilitates transparent, reproducible UQ analysis for both the development and application of IPs in molecular modeling.
We focus on a Bayesian, MCMC approach for quantifying parametric uncertainty and model inadequacy.
We use a parallel-tempered MCMC method, which improves convergence and allows us to mimic the effect of model inadequacy in our uncertainty estimates.

In our implementation we focused on ease of use in order to lower the barrier to entry for molecular modeling practitioners.
However, we caution users not to treat these methods as off-the-shelf black boxes.
UQ is an emerging field with many open questions, especially surrounding model inadequacy which is often the dominant problem in atomistic simulations.
We encourage IP practitioners and developers alike to familiarize themselves with statistical subtleties related to sloppiness and parameter unidentifiability\cite{whynonlinearfits, kurniawan2021bayesian} and check their conclusions for robustness over a range of sampling temperatures for multiple priors.

% Future work
In future work, we plan to integrate other UQ methods within KLIFF, such as the frequentist profile likelihood\cite{Cole_Chu_Greenland_2014}.
To address the problems associated with UQ of sloppy IPs\cite{kurniawan2021bayesian}, we plan to integrate a model reduction scheme motivated by information geometry\cite{MBAM}.

\section{Acknowledgment}
This work is supported by the National Science Foundation under awards DMR-1834332 and DMR-1834251.
We would like to acknowledge the computational facilities provided by the Brigham Young University Office of Research Computing.

\bibliographystyle{IEEEtran}
\bibliography{refs.bib}

\end{document}